\documentclass[seceq]{ptptex}

\usepackage{graphicx}
\usepackage{latexsym}
\usepackage{wrapft}

\title{
Competing Orders and non-Landau-Ginzburg-Wilson Criticality in (Bose) Mott transitions%
}

\inst{
$^1$Physics Department, Tokyo Bunrika University, Tokyo 113-1234, Japan
\\
$^2$Yukawa Institute for Theoretical Physics, Kyoto University, \\
Kyoto 606-8502, Japan}

\author{
  Leon \textsc{Balents}$^{1}$, Lorenz \textsc{Bartosch}$^{2,3}$, Anton
  \textsc{Burkov}$^1$, Subir \textsc{Sachdev}$^2$, and Krishnendu
  \textsc{Sengupta}$^2$}

\inst{
$^1$Physics Department, University of California, Santa Barbara, \\
$^2$Department of Physics, Yale University, P.O. Box
208120, New Haven, CT 06520-8120 \\
$^3$Institut f\"ur Theoretische Physik, Universit\"at
Frankfurt, Postfach 111932, 60054 Frankfurt, Germany}

\abst{
  This paper reviews a recent non-Landau-Ginzburg-Wilson (LGW)
  approach to superfluid to Mott insulator transitions in two
  dimensional bosonic lattice systems, using a dual vortex field
  theory.\cite{psgbosons,psgdimers} The physical interpretation of
  conventional LGW theory of quantum criticality is re-examined and
  similarities and differences with the vortex picture are discussed.  The
  ``unification'' of various competing (insulating) orders, and the
  coincidence of these orders with the Mott transition are readily
  understood in this formulation.  Some aspects of the recent theory
  of ``deconfined'' quantum criticality, which are to an extent
  subsumed in this approach, are discussed.  A pedagogical
  presentation of the ``nuts and bolts'' of boson-vortex duality at
  the hamiltonian level is included, tailored to a condensed matter
  audience. }

\begin{document}

\maketitle

\section{Introduction}
\label{sec:introduction}

Condensed matter theory continues to be challenged by the remarkable
behavior of strongly correlated materials.  The archetype of strong
correlation physics is the Mott insulator, in which insulating behavior
arises not due to the presence of filled bands (ultimately the Pauli
exclusion principle), but to charge localization by strong local Coulomb
interactions.  Locally this physics is extremely simple, but extending
it away from the atomic context, to the many-body problem is not so
straightforward.  The theoretical treatment of Mott insulators -- and
systems proximate to a Mott state -- is complicated by the fact that
such Mott localization reflects neither a feature of some electron-like
quasiparticle spectrum, nor any kind of symmetry breaking.  Thus the two
workhorses of solid state physics, Landau's Fermi liquid theory and
the symmetry/order parameter description of phases of matter, are not
helpful in describing the basic Mott physics.  

In practice, most Mott insulators order at low temperatures, either
magnetically, or by charge-order (charge density wave, stripe, etc.).
These various orderings can be viewed as ``competing'' with one
another, and with electronically-extended phases in which the Mott
localization physics is not operative.  Such competing orders are
increasingly observable at the microscale through improvements in STM
microscopy,\cite{krmp}\tocite{hanaguri} and in crystal and surface
quality.  A complication for theoretical treatments of competing
orders is that there are too many different charge and/or spin
ordering patterns, often of considerable complexity, consistent with
the simple local Mott physics requiring only single occupancy of some
orbitals.  Therefore we view these orders as derivative phenomena,
occuring as a consequence of Mott localization, not vice-versa.

Nevertheless, competing orders seem to be an essential accompaniment
to Mott localization.  Indeed, there have been a number of ``proofs''
of late (some of which may even be rigorous!) that seem to require
order of some sort in a Mott state which does not have an average
electron occupation which is even per unit
cell.\cite{Hastings,Oshikawa} \ More specifically, these proofs
require ground state degeneracies in large systems, which may be
associated physically either to broken symmetry (presumably the usual
case) or to more exotic ``topological order''.  Thus a minimal
requirement of any reasonable theory of the Mott transition is that it
should lead automatically to conventional or at least topological
order.

In this paper, we discuss the SuperFluid (SF) to Mott transition in
two-dimensional lattice {\sl boson} systems.\cite{FWGF} This problem has
recently become a very active experimental area with the advent of cold
trapped atoms in an optical lattice.\cite{Greiner} To connect it
theoretically to the electronic Mott physics discussed above, one can
view the bosons as tightly-bound ``Cooper pairs''; the above
arguments for the necessity of competing order in the insulator thus
apply when the boson filling non-integral.  We give a somewhat
pedagogical review of recent work\cite{psgbosons,psgdimers}\ which
satisfies the above requirement of naturally describing the emergence of
competing orders in the insulator, while not putting this in ``by
hand''.  We focus on rational fillings $f=p/q$, with $p,q$ relatively
prime, and $q>1$.  This is accomplished by the use of {\sl
  duality},\cite{dh,nelson,fisherlee} a technical transformation of the
hamiltonian which has the utility of allowing one to approach the
SF-Mott (putative) quantum critical point from the {\sl superfluid}
side, the opposite of the Landau-Ginzburg-Wilson (LGW) approach to
superfluidity,\cite{FWGF} which expands about the normal phase in terms of the
superfluid order parameter.  We argue that, because of the order in the
insulator, such an LGW approach is unnatural.

Thinking of the Mott transition as a Quantum Critical Point (QCP),
understanding it is just the search for an appropriate continuum
quantum field theory.  Quantum field theories are ``second quantized''
descriptions of particles: point-like excitations created and
annihilated by the quantum fields.  LGW theory takes, loosely
speaking, these particles to be the original lattice bosons.  The dual
approach described in this paper uses instead the {\sl vortices} of
the superfluid as these ``particles''.  Because it is formulated in
terms of vortex excitations, this approach 
takes advantage of the fact that the nature of the superfluid ground
state (in particular its symmetries) is {\sl independent} of the boson
filling.  

What begins as a worry -- the non-local nature of the phase
gradient/superflow surrounding a vortex -- ends up as an advantage in
the approach.  In the vortex formulation, this non-locality is
completely accounted for by a dual gauge field $A_\mu$, which couples
minimally to the vortices.  The gauge field accounts for the phase
winding of the boson wavefunction upon encircling a vortex.  Turned
around, the vortex wavefunction must wind when encircling a boson.
The phase winding due to the average boson number encircled by a
minimal motion of a vortex on the lattice, is captured in the dual
theory by an Aharonov-Bohm ``flux'' of $2\pi f$ in $A_\mu$ through a
dual plaquette.\footnote{The same approach has been expored in
  Refs.~\citen{lfs,sp,tesanovic}.} When the bosons are at non-integer
filling, this modifies the low-energy vortex dynamics in an essential
way.  In particular, it forces the vortices to appear in {\sl
  multiplets} of $q$ flavors, described by a vector of vortex fields,
$\varphi_\ell$, $\ell=0\ldots q-1$.  These multiplets transform under
a {\sl projective representation} of the physical lattice symmetry
group, a generalization of the ordinary notion of a representation, in
which the group multiplication table is obeyed only up to phase
factors, i.e. if a phase is ``projected out''.  An important example
is the $x$ and $y$ lattice translations on the square lattice, which
obey
\begin{equation}
  T_x T_y = \omega T_y T_x , \label{e7a}
\end{equation}
with $\omega=e^{2\pi i f}$.  This mathematically captures the
Aharonov-Bohm phase described above.  The modified group is dubbed a
Projective Symmetry Group, or PSG.\cite{wenpsg} A PSG is possible only
because of the gauge nature of the vortex theory, ``projection'' being
allowed by the lack of independent physical meaning of the local phase
of the vortex field.

The PSG dictates the form of the dual effective action, similarly to 
how the ordinary symmetry group dictates the free energy in LGW theory.
One finds (see Sec.~\ref{sec:non-integral-mott})
\begin{equation}
  \label{eq:Sq}
  \mathcal{S} =  \int d^2 r d \tau \Big\{ \frac{1}{2e^2} \left( \epsilon_{\mu\nu\lambda}
\partial_\nu A_\lambda \right)^2 + \sum_\ell
\left[ |(\partial_\mu - i A_{\mu} ) \varphi_\ell |^2 + \tilde{r}
|\varphi_\ell |^2 \right] + {\cal L}_{\rm int} \Big\},
\end{equation}
where ${\cal L}_{\rm int}$ represents quartic and higher order terms
in the $\{\varphi_\ell\}$.  , which are strongly constrained by the PSG.
The form of ${\cal L}_{\rm int}$ is, however, specific to each value
of $q$ (see Ref.~\citen{psgbosons} for a general discussion, and
Sec.~\ref{sec:examples} for examples).

Eq.~(\ref{eq:Sq}) describes the superfluid/Mott physics through the
gauge field $A_\mu$.  For instance, the phason mode of the superfluid
is described in Eq.~(\ref{eq:Sq}) as the gapless transverse ``photon''
mode on the gauge field.  The condensation of any of the
$\varphi_\ell$ fields leads to a ``Higgs'' mass for the gauge field,
corresponding to the loss of the photon and the gap in the Mott
phase.  Competing ``charge'' orders are less apparent in
Eq.~(\ref{eq:Sq}), but they are in fact encoded in the structure of
the PSG.

In particular, the order parameters for the different charge ordering
patterns occuring in Mott states can be written in terms of ``density
wave'' amplitudes, $\rho_{\bf Q}$, describing the (complex) amplitude
of a plane-wave oscillation in the charge density at wavevector ${\bf
  Q}$.  The non-trivial vortex PSG provides a link between the
vortex multiplet and spatial symmetry operations.  In fact, one can
explicitly construct a set of such density wave operators, with
\begin{equation}
  {\bf Q}_{mn} = 2 \pi f (m,n), \label{e10}
\end{equation}
where $m,n$ are integers.  In particular, we find (see
Sec.~\ref{sec:order-parameters}) 
\begin{equation}
  \rho_{mn} \equiv \rho_{{\bf Q}_{mn}} = S\left(|{\bf Q}_{mn}|\right
  ) \omega^{mn/2} \sum_{\ell = 0}^{q-1} \varphi^{\ast}_\ell
  \varphi_{\ell+n} \omega^{\ell m}. \label{e11}
\end{equation}
In associating the $\rho_{mn}$ with a density, there is a general
`form-factor', $S(Q)$, which cannot be determined
from symmetry considerations, and has a smooth $Q$ dependence
determined by microscopic details and the precise definition of
the density operator.  

Unlike in Landau theory, the density-wave order parameters describing
the possible ``competing orders'' in the Mott state are quadratic
rather than linear in the quantum fields of the theory.  The ordering
is thus ``weaker'' than might be expected near a LGW-type charge
ordering transition, consistent with the notion that the vortex action
describes the Mott transition first, and competing charge orders in
the insulator only as a secondary consequence.  Describing both
phenomena -- the loss of superfluidity and onset of charge order --
simultaneously is already an achievement.  It is possible because,
since the $\varphi_\ell$ field carries the dual gauge charge, vortex
condensation can describe both Mott physics (through the Higgs mass of
$A_\mu$ when $\langle\varphi_\ell\rangle\neq 0$) and the emergent
competing order arising through the $\rho_{mn}$.  An intriguing
conseqence is that, even in the superfluid phase, one expects to see
density wave modulations appearing in the vicinity of a localized
vortex.\cite{psgbosons}  Equivalent but complementary viewpoints of
this non-LGW quantum criticality for the special case of half-filling
have been extensively developed in recent work on ``deconfined quantum
criticality''\cite{dqcp} (see also Sec.~\ref{sec:deconf-crit}).  

This paper is written to provide a ``gentler'' introduction to the
work of Ref.~\citen{psgbosons}.  To keep it pedagogical, it was not
possible to go beyond this work to discuss the intended {\sl applications} of
these theoretical ideas to {\sl electronic} systems near a {\sl
  superconductor} to Mott insulator transition, with an eye to the
under-doped cuprate materials.  We cannot resist, however, pointing
the reader toward this interesting direction.  In
Ref.~\citen{psgdimers}, it was demonstrated that the same dual critical
field theory applies to a somewhat more microscopically faithful
representation of strong electronic pairing, the doped quantum dimer
model.\cite{fradkiv} The authors believe that it indeed applies more
generally to any two-dimensional {\sl clean} singlet superconductor to
Mott insulator transition, in which the gap in the superconducting
state is complete.  Most recently, it was proposed that the
observation of ``checkerboard'' charge correlations near vortices in
BSCCO by Hoffman {\sl et al}\cite{Hoffman}, is indicative of similar
physics.  This suggestion was used to extract some bounds on the
inertial mass of a vortex from these experiments.\cite{mass} \ The
reader should note that, because the cuprate materials are gapless,
d-wave superconductors, this interpretation goes boldly beyond the
existing theory.  Theoretical work to directly address the additional
quasiparticle physics in gapless superconductors is ongoing.

The remainder of this paper is organized as follows.  In
Sec.~\ref{sec:bose-mott-insulators}, we describe the general structure of
simple models of lattice bosons, which contain superfluid to Mott
transitions.  Sec.~\ref{sec:integr-mott-superfl} describes the
conventional LGW theory of the SF-Mott QCP for {\sl integer} boson
filling, emphasizing that it should be understood as based upon
elementary particle/hole excitations of the Mott state.
Sec.~\ref{sec:integr-mott-superfl-1} describes the dual vortex
description of this conventional integral filling Mott transition,
including a pedagogical description of hamiltonian boson-vortex
duality.  Sec.~\ref{sec:non-integral-mott} applies the dual
formulation to non-integral Mott transitions, and describes the origin
of the main results summarized in this introduction.  Finally,
Sec.~\ref{sec:examples} details some specific predictions of the
theory for $f=1/2$, which provides an example and view to ``deconfined
criticality'', and for $f=1/3$, which does not.

\section{Bose Mott Insulators}
\label{sec:bose-mott-insulators}

\subsection{Physics and model}
\label{sec:physics-model}

The simplest systems that can exhibit a Mott transition between states
with extended and localized carriers are those composed of interacting
lattice bosons, the extended states being superfluids/superconductors,
and the Mott states being generally charge ordered away from integer
boson fillings.  An exciting experimental development of late is the
direct realization of such models of cold atomic bosons confined to an
optical lattice.  These will likely turn out to provide the cleanest
experimental tests of theoretical approaches to this simplest Mott
problem.  Conceptually, a symmetry-equivalent problem arises if fermions
(e.g. electrons) are strongly bound into bosonic Cooper pairs.  This is
not clearly realized in any simple electronic material.  One may expect,
however, that a bosonic theory of this sort will properly describe the
{\sl universal} low energy properties of an electronic system in the
vicinity of a superconductor to Mott insulator transition, provided that
both the superconductor and insulator are ``bosonic''.  More
precisely, we believe it clearly applies when
the superconductor should be clean (strictly speaking, {\sl superclean})
and with a full gap (e.g. s-wave) to unpaired quasiparticle excitations,
and the Mott insulator should exhibit only charge (e.g. not spin)
order.  Similar theories likely apply more generally to other superconductor
to Mott insulator transitions, but are beyond the scope of this paper.

We concentrate now on purely bosonic Mott transitions.  We use a
``rotor'' representation for the bosons, with phase operators
$\hat{\phi}_i$ and conjugate number operators $\hat{n}_i$ where $i$
runs over the sites of the direct lattice.  These operators obey the
commutation relation
\begin{equation}
[\hat{\phi}_i, \hat{n}_j] = i \delta_{ij}.
\end{equation} 
One may consider a variety of geometries, but we will focus here upon
the simple square lattice.

A simple boson hamiltonian in the class of interest has the
structure
\begin{equation}
\mathcal{H} = -t \sum_{i\alpha} \cos \left(\Delta_{\alpha}
\hat{\phi}_i \right) + U \sum_i (\hat{n}_i-\overline{n})^2  + \mathcal{H}',
\label{hubbard} 
\end{equation}
where $t$ represents a nearest-neighbor boson hopping amplitude, and
$U$ an on-site boson repulsion.  We denote by $\Delta_\alpha$ the
discrete lattice gradient in the $\alpha$ direction.  Additional more
complicated off-site terms are included in the unspecified
contribution to the hamiltonian, $\mathcal{H}'$ (but see below).  For
now we require only that it respect the symmetries of the underlying
lattice, boson conservation, and locality (e.g. the amplitude for
hopping between far away sites should decay sufficiently rapidly with
distance).

Eq.~(\ref{hubbard}) may be studied in the grand canonical ensemble,
i.e. at fixed chemical potential ($2U\overline{n}$).  However, one may
equally well consider instead the canonical ensemble with fixed boson
number (average filling $f$).  We will typically do the latter, except
in Secs.~\ref{sec:mott-states-at},\ref{sec:mott-states-at-1}, and the
first part of Sec.~\ref{sec:integr-mott-superfl}, where we work at
fixed chemical potential.

\subsection{Mott states at integral filling}
\label{sec:mott-states-at}

\begin{wrapfigure}{r}{6.6cm}
\centerline{\includegraphics[width=4.0cm]{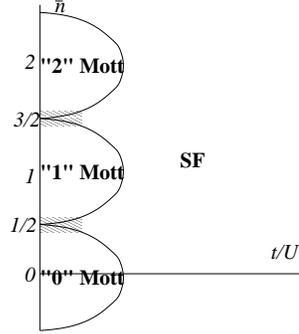}}
\caption{Schematic phase diagram of boson rotor model, with on-site
  interactions only.  The shaded regions indicate where a there is a
  large near-degeneracy of states with different boson densities, and
  the system is highly susceptible to off-site interactions.}
\label{fig:1}
\end{wrapfigure}

Neglecting terms in $\mathcal{H}'$, the zero temperature phase diagram
of $\mathcal{H}$ is well-known.\cite{FWGF} It takes the schematic form
in Fig.~\ref{fig:1}.  For $t/U \ll 1$, the system is in a Mott
insulating ground state, with $\langle \hat{n}_i\rangle = N$, the
integer nearest to $\overline{n}$, on every site.  This phase persists
inside the ``lobes'' drawn in the figure.  There is a gap to the
lowest-lying excited states, which may be thought of as single
extra/missing bosons (which delocalize into plane-waves).  For large
$t/U$, the ground state is a superfluid ({\bf SF} in the figure), with
$\langle e^{i\hat{\phi}_i}\rangle =\Psi_{sf} \neq 0$, and the density
$f=\langle \hat{n}_i\rangle$ varies smoothly with parameters in an
unquantized fashion.  There is no excitation gap, and the lowest-lying
excitations are acoustic ``phonons'' or ``phasons'', the Goldstone
modes of the broken U(1) symmetry of the superfluid.

\subsection{Mott states at non-integral filling}
\label{sec:mott-states-at-1}

We now return to the shaded regions of the phase diagram in
Fig.~\ref{fig:1}, where states with different boson density are nearly
degenerate.  Indeed, in the simple model with $\mathcal{H}'=0$, for
$\overline{n}=N+1/2$, states with any average density between $N$ and
$N+1$ are degenerate.  For $t/U=0$, the eigenvalue of $\hat{n}_i=N$ or
$\hat{n}_i=N+1$ can be independently chosen on each site.  The omitted
terms in $\mathcal{H}'$ will then clearly determine the nature of the
ground states appearing in the shaded region.  Generally, Mott
insulating states appear at rational fractional fillings, $f=p/q$,
with $p,q$ relatively prime.  For $q> 1$, these are boson ``crystals''
or charge density waves.  Mott states with increasing $q$ are expected
to require longer-range interactions in $\mathcal{H}'$ for their
stabilization.
\begin{wrapfigure}[10]{r}{6.6cm}
\centerline{\includegraphics[width=4.0cm]{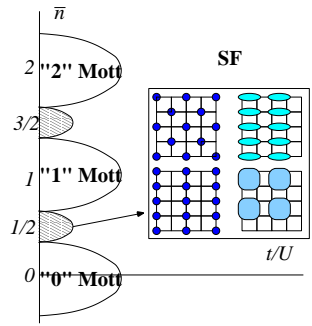}}
\caption{Schematic phase diagram with off-site interactions.  Some
  representative Mott insulating states with $\langle
  \hat{n}_i\rangle=N+1/2$ are shown.}
\label{fig:2}
\end{wrapfigure}

For example, in the vicinity of $\overline{n}=N+1/2$, we can adopt a
pseudo-spin description, with $S_i^z = \hat{n}_i-N-1/2 = \pm 1/2$.  In
the limit of $U\rightarrow \infty$ (or $t/U\ll 1$), one can then
replace
\begin{equation}
  \label{eq:HUinf}
  \mathcal{H} \rightarrow -t \sum_{\langle ij\rangle} \left( S_i^+
    S_j^- + S_i^- S_j^+\right),
\end{equation}
with $\langle ij\rangle$ indicating the sum is taken over
nearest-neighbor sites, and the conventional definitions of spin-$1/2$
raising and lowering operators.  This term tends to split the
degeneracy of boson states in favor of a delocalized superfluid.
Terms in $\mathcal{H}'$ can compete with this term for small $t$,
splitting the degeneracy instead in favor of a non-trivial Mott state.
For instance, one may take
\begin{equation}
  \label{eq:Hpinf}
  \mathcal{H}' = J_z \sum_{\langle ij\rangle} S_i^z S_j^z + J'_z
  \sum_{\langle\langle ij\rangle\rangle}S_i^z S_j^z  - K \sum_{\Box}
  \left( S_i^+ S_j^- S_k^+ S_l^- + {\rm h.c.}\right)+\cdots
\end{equation}
Here $\langle\langle ij\rangle\rangle$ indicates a sum over
next-nearest-neighbors, and the $\Box$ a sum over four site
plaquettes (with sites $i,j,k,l$ labelled clockwise around the
plaquette).  For $t \ll \min \{ J_z, J'_z, K\}$, a variety of Mott
states occur in such a model, including ``solids'' and ``valence bond
solids'' as shown in Fig.~\ref{fig:2}.  The figure also shows a
schematic phase diagram in which Mott states with $q=2$ are stable.
More complex phases (with $q>2$) have been observed in similar models
in the literature.\cite{otterlo}

\section{``Integral'' Mott-Superfluid transitions: LGW theory as Bose
  condensation}
\label{sec:integr-mott-superfl}

It is well-known that the quantum phase transitions across the phase
boundaries in Fig.~\ref{fig:1} {\sl are} described by LGW theories.\cite{FWGF}
On symmetry grounds, this is understandable, since the assumptions
underlying Landau's theory are valid.  In particular, the integer
filling Mott states, being unique, and having the full symmetry of the
underlying hamiltonian and only short-range correlations, may be
regarded as truly ``disordered''.  The superfluid is the ``ordered''
phase, its symmetry group (the lattice space group) being a subgroup
of the symmetry group of the Mott state (the direct product of the
lattice space group and the $U(1)$ boson number conservation
symmetry).  A LGW-style expansion of the effective action in terms of
the superfluid order parameter $\Psi$ and its derivatives indeed seems
to describe the quantum critical points.

It is instructive, however, to understand these transitions more
physically.  We will demonstrate that the physical content of the LGW
theory is Bose-Einstein condensation of ``particle'' and/or ``hole''
excitations of the Mott state.  To that end, let us think about the
elementary excitations in that phase.  For $U/t \gg 1$, the ground
state is simply
\begin{equation}
  \label{eq:MGS}
  |GS\rangle = \prod_i (b_i^\dagger)^N |0\rangle,
\end{equation}
where $b_i=e^{-i\hat\phi_i}$ is the boson annihilation operator in the
rotor formulation, and $|0\rangle$ is the state of ``no bosons'' (zero
rotor angular momentum), $\hat{n}_i|0\rangle=0$ (note that eigenstates
of $\hat{n}_i$ exist with negative as well as non-negative integer
eigenvalues in the rotor formulation).  In the same limit, the lowest
excited states are ``particles'' and ``holes'' with one extra or
missing boson,
\begin{eqnarray}
  \label{eq:Mex}
  |p_i\rangle & = & b_i^\dagger |GS\rangle, \\
  |h_i\rangle & = & b_i |GS\rangle.
\end{eqnarray}
For $t/U=0$ strictly, the particle and hole energies (relative to the
ground state) are
\begin{equation}
  \label{eq:Eph}
  E^{(0)}_p = 2 U (N+ \frac{1}{2}-\overline{n}), \qquad E^{(0)}_h = 2
  U (\overline{n}+\frac{1}{2}-N). 
\end{equation}
For $0<t/U \ll 1$, these states will develop dispersion.  By
considering the first order splitting of the degenerate manifold of
particle or hole states (degeneracy associated with the site of the
particle or hole), one obtains
\begin{equation}
  \label{eq:Epht}
  E_{p/h} = E^{(0)}_{p/h} - 2 t (\cos k_x + \cos k_y)\approx
  \Delta_{p/h} + \frac{k^2}{2m},
\end{equation}
where we have Taylor expanded around the minimum at ${\bf k=0}$,
giving $m= 1/(2t)$.  
The minimum energy excitation is then a particle or hole, for $\overline{n}>N$
or $\overline{n}<N$, respectively.
The excitation gap, $\Delta_{p/h}$, is
\begin{equation}
  \label{eq:Egap}
  \Delta_{p/h} = 2U(\frac{1}{2} \mp (\overline{n}-N)) - 4t,
\end{equation}
to this order.  A simple-minded extrapolation of this formula to
larger $t$ suggests that the excitation gap for particles or holes
will vanish once $t/U \geq (\frac{1}{2} -|\overline{n}-N|)/2$.  Though
this $O(1)$ value is well beyond the region of validity of the
expansion (except for $|\overline{n}-N|\approx \frac{1}{2}$, where it
is invalid for other reasons, to be discussed below), it does suggest
a simple physical picture of the transition to the superfluid, as a
``condensation'' of these particles or holes.  For $\overline{n}>N$,
the particles condense, while for $\overline{n}<N$ the holes do.
Precisely at $\overline{n}=N$, both condense simultaneously.  This
latter case automatically applies in the canonical ensemble if the
filling $f$ is fixed to be integral.

We will now show that such particle/hole condensation
{\sl is} the physical content of the LGW theory.
While it is possible to derive a field theory of this condensation
from $\mathcal{H}$, we instead just write it down based on this simple
physical picture.  We model the particle and hole excitations by
fields $p({\bf x},\tau), h({\bf x},\tau)$ respectively, in the
imaginary time ($\tau$) path integral.  The weight in the path
integral is, as usual, the Euclidean action,
\begin{equation}
  \label{eq:Sph}
   \mathcal{S} = \int\!d\tau\!d^2x\, \big[ p^\dagger(\partial_\tau +
  E_p - \frac{1}{2m}\nabla^2)p^{\vphantom\dagger} + h^\dagger(\partial_\tau +
  E_h - \frac{1}{2m}\nabla^2)h^{\vphantom\dagger} - \lambda(p^\dagger
  h^\dagger + p^{\vphantom\dagger}h^{\vphantom\dagger}) + \cdots\big].
\end{equation}
Here we have included a term $\lambda$ which creates and annihilates
particles and holes together in pairs, which is expected since this
conserves boson number.  Microscopically this term arises from the
action of the hopping $t$ on the na\"ive ground state, which creates
particle-hole pairs on neighboring sites, so $\lambda \sim O(t)$ (the
spatial dependence is unimportant for the states near ${\bf k=0}$).
We have neglected -- for brevity of presentation -- to write a number
of higher order terms involving four or more boson fields,
representing interactions between particles and/or holes, and other
boson number-conserving two-body and higher-body collisional
processes.  Note that the dependence upon $t/U$ in Eq.~(\ref{eq:Sph})
arises primarily through implicit dependence of $E_{p/h}$.

Eq.~(\ref{eq:Sph}) can be systematically analyzed by diagonalizing the
quadratic form involving $p,h$.  The qualitative features can be
understood even more simply by considering simple limits.  First,
suppose $\overline{n}>N$, so $E_p<E_h$.  Upon increasing $t/U$, then, at some
point $E_p \rightarrow 0$ while $E_h>0$.  Then the holes are still
gapped, and the $h$ field can be integrated out order by order in
$\lambda$ and higher order terms, encountering no vanishing energy
denominators.  One has then an effective action for the remaining
particle field $p$,
\begin{equation}
  \label{eq:Sp}
  \mathcal{S} = \int\!d\tau\!d^2x\, \big[ p^\dagger(\partial_\tau +
  \tilde{E}_p - \frac{1}{2m}\nabla^2)p^{\vphantom\dagger} + u
  (p^\dagger p^{\vphantom\dagger})^2 +\cdots\big],
\end{equation}
with $\tilde{E}_p \approx E_p - O(\lambda^2/E_h)$ a renormalized
particle gap.  This is exactly the well-known effective action for the
superfluid-Mott transition, which has an LGW form with the ``order
parameter'' being the particle field $p$.  The single $\partial_\tau$
derivative is well-understood and present because particle
annihilation becomes particle creation under time-reversal, and leads
to so-called $z=2$ behavior of the QCP.  

The case of $\overline{n}<N$ can be understood similarly with the roles of
particles and holes interchanged.  The remaining case of $\overline{n}=N$ (the
tips of the lobes in Fig.~\ref{fig:1}) is slightly different.  In that
case, both particles and holes become gapless together as $t/U$ is
increased.  Thus we may expect a ``relativistic'' theory in which
particles and holes appear in a sense as particles and antiparticles.
This is seen by changing variables to the linear combinations
\begin{equation}
  \label{eq:lc}
  \Psi = \frac{1}{\sqrt{2}}(p^{\vphantom\dagger}+h^\dagger) \qquad
  \Xi = \frac{1}{\sqrt{2}}(p^{\vphantom\dagger}-h^\dagger). 
\end{equation}
Writing $E_p=E_h\equiv E_0$, one finds that (taking $\lambda>0$
without loss of generality) the quadratic form for $\Psi$ becomes
unstable before that of $\Xi$, and one can therefore integrate out
$\Xi$ to obtain
\begin{equation}
  \label{eq:Spsi}
  \mathcal{S} = \int\!d\tau\!d^2x\, \big[
  \Psi^\dagger(-\partial_\tau^2  -
  \frac{1}{2m}\nabla^2+r)\Psi^{\vphantom\dagger} + u 
  (\Psi^\dagger \Psi^{\vphantom\dagger})^2 +\cdots\big],
\end{equation}
with $r \approx E_0-\lambda$.  Apart from a trivial
space-time anisotropy, Eq.~(\ref{eq:Spsi}) is exactly the classical
LGW free energy of a three-dimensional normal-superfluid transition.
Thus the two LGW actions, Eqs.~(\ref{eq:Sp},\ref{eq:Spsi}), usually denoted
as particle-hole asymmetric and symmetric theories, respectively, are
indeed physically equivalent to particle or particle+hole
condensation.

\section{``Integral'' Mott-Superfluid transitions: vortex condensation
  theory}
\label{sec:integr-mott-superfl-1}

As discussed in the introduction, a quantum field theory
description of the QCP would appear to require only some pointlike
``particles'' described by the quantum fields (as
creation/annihilation operators).  On the Mott side of the transition,
particle and/or hole excitations provide these, and their condensation
precisely coincides with LGW theory.  On the superfluid side, there is
another quite different pointlike ``particle'' excitation: a vortex or
anti-vortex.  This is specific to two-dimensional superfluids, since a vortex
becomes a line defect in three dimensions.  

It is natural then to try to describe the Superfluid-Mott transition,
coming from the superfluid side by a field theory for vortices.
Provided time-reversal symmetry $\mathcal{T}$ is unbroken, one should
expect such a field theory to be relativistic, since vortices and
anti-vortices are interchanged by $\mathcal{T}$ (note that this is
independent of the presence or absence of particle/hole symmetry in the
boson system).  One may worry that a vortex is a non-local object, with
a power-law tail of superflow surrounding it extending to infinity.
Perhaps this invalidates its use as a particle?

\subsection{Duality}
\label{sec:duality}

This worry is resolved by {\sl duality}, which is a rigorous
mathematical mapping of the original rotor boson model to one of
vortices.  We will see that all the non-locality of the vortex is taken
into account by a {\sl non-compact $U(1)$ gauge field}.  The dual
formulation is mathematically analogous to a lattice $U(1)$ Higgs theory of
particle physics, or a lattice classical three-dimensional
Ginzburg-Landau theory for a superconductor (charged superfluid).

Boson-vortex duality is extensively covered in the literature.  It is
however often a confusing topic for students and senior researchers
alike.  This is because certain stages of the duality mapping are
usually performed in a rather non-rigorous fashion, involving so-called
``Villain potentials'', fudging with the time-continuum limit in the
path integral, and other similar manipulations.  In reality this
sloppiness is unimportant, since the usefulness of duality does not lie
in the quantitative analysis of specific microscopic models, but in its
application to universal phenomena.  Nevertheless, it seems anathema to
those trained in the more rigorous solid state physics tradition rather
than the effective field theory approach originating from statistical
mechanics.  

For this reason we will present here what to our mind is the simplest
and most rigorous possible version of $U(1)$ duality, performed at the
hamiltonian level.  This proceeds in several steps.  The first,
completely rigorous and explicit step, contains the essence of the
duality mapping.  It is simply a change of variables from boson number
and phase $\hat{n}_i, \hat\phi_i$ living on the direct lattice sites, to
new ``electric field'' and ``vector potential'' variables, $E_{a\alpha},
A_{a\alpha}$, living on the links of the dual lattice (We use $a,b,c$ to
label sites of the dual lattice, and $\alpha,\beta=1,2$ to label links of the
dual square lattice in the $x,y$ directions, respectively):
\begin{eqnarray}
  \label{eq:Edef}
  E_{a\alpha} & = & \frac{(\epsilon_{\alpha\beta}\Delta_\beta \hat\phi)_a}{2\pi}, \\
  \hat{n}_i & = & \frac{(\epsilon_{\alpha\beta}\Delta_\alpha A_{\beta})_i}{2\pi}.
  \label{eq:Adef} 
\end{eqnarray}
Geometrically, the oriented electric field on a link of the dual lattice
is taken to equal $1/(2\pi)$ times the phase difference between the
phase $\hat\phi_i$ to immediately to the right (using the orientation of
the field) of the dual link and the phase $\hat\phi_j$ immediately to
the left of this dual link.  The electric field $E_{a\alpha}$ is thus a
periodic variable, with $E_{a\alpha} \leftrightarrow E_{a\alpha}+1$.  The dual
vector potential $A_{a\alpha}$ is a discrete field, i.e. has eigenvalues of
$2\pi$ times integers.  It is implicitly defined by Eq.~(\ref{eq:Adef})
so that its lattice curl -- the counter-clockwise circulation around a
dual plaquette -- is $2\pi$ times the boson number inside this dual
plaquette.  The $A_{a\alpha}$ variables have a ``gauge'' redundancy: the
vector potential can be shifted by any $2\pi\times$integer gradient,
$A_{a\alpha}\rightarrow A_{a\alpha}+\Delta_\alpha \chi_a$ (with $\chi_a\in
2\pi\mathcal{Z}$) without changing $\hat{n}_i$.  To achieve a one-to-one
mapping of $A_{a\alpha}$ to $\hat{n}_i$, one can ``fix a gauge'' in
numerous ways, e.g. for Coulomb gauge, $\Delta_\alpha A_{a\alpha}=0$.

It is straightforward to check (by comparing the commutator of the
left-hand side of Eq.~(\ref{eq:Edef}) and the right-hand side of
Eq.~(\ref{eq:Adef}) with the commutator of the right-hand side of
Eq.~(\ref{eq:Edef}) and the left-hand side of Eq.~(\ref{eq:Adef})), that
the dual electric and vector potential variables are
canonically-conjugate:
\begin{equation}
  \label{eq:EAcom}
  \left[ A_{a\alpha}, E_{b\beta}\right] = i \delta_{ab}\delta_{\alpha\beta}.
\end{equation}
Furthermore, the expression in Eq.~(\ref{eq:Edef}) implies a constraint:
\begin{equation}
  \label{eq:Gauss1}
  \Delta_\alpha E_{a\alpha} \in \mathcal{Z},
\end{equation}
i.e. the lattice divergence of the dual electric field is an integer (a
non-zero value being ``allowed'' due to the periodicity of the
$\hat\phi_i$ and $E_{a\alpha}$ variables).  Physically, this integer can
be identified with the vorticity.  This can be seen by considering the
line sum of $\Delta_\alpha \hat\phi_i$ around some area.  In an abuse
of the continuum notation,
\begin{equation}
  \label{eq:vorticity}
  \oint {\boldsymbol{\nabla}}\hat\phi\cdot d\boldsymbol{l} = 2\pi \oint
  {\boldsymbol{E}}\cdot d\boldsymbol{\hat{n}}=2\pi \int \! d^2x \,
  \Delta_\alpha E_{a\alpha},
\end{equation}
so that the phase winding around some area, in units of $2\pi$, just
counts the total divergence of $E_{a\alpha}$ inside this area, i.e.
the net dual ``charge'' inside this area.

With full rigor,\footnote{Some small additional care needs to be taken
  for finite systems with periodic boundary conditions, in which case
  additional c-number terms should be added to the left-hand side of
  Eq.~(\ref{eq:Adef}).} the hamiltonian can be rewritten as
\begin{equation}
  \label{eq:Hdual1}
  \mathcal{H} = -t \sum_{a\alpha} \cos 2\pi E_{a\alpha} + \frac{U}{(2\pi)^2}
  \sum_i \left( (\epsilon_{\alpha\beta}\Delta_\alpha A_{\beta})_i-2\pi \overline{n}\right)^2.
\end{equation}

Already without further manipulation, Eq.~(\ref{eq:Hdual1}) appears very
similar to an electromagnetic hamiltonian ($\propto E^2 +B^2$).  It is,
however, inconvenient to work with since $A_{a\alpha}$ is a discrete,
$2\pi\times$integer-valued field.  To understand how to remedy this
deficiency, it is instructive to express the first term in
Eq.~(\ref{eq:Hdual1}) in the $A_{a\alpha}$ basis.  Since the electric field
is conjugate to the vector potential, this cosine is just a ``shift''
operator for the $A_{a\alpha}$ field.  Hence we can write
\begin{equation}
  \label{eq:Hdual2}
  -t \cos 2\pi E_{a\alpha} = -t \sum_{A_{a\alpha}} \big(
    |A_{a\alpha}+2\pi\rangle\langle A_{a\alpha}|+
    |A_{a\alpha}\rangle\langle A_{a\alpha}+2\pi|  \big).
\end{equation}
One can think of the basis for the Hilbert space of a single link of the
dual lattice as consisting of $2\pi\times$integer-spaced ``sites'' along
a ``line'' $A_{a\alpha}$-space.  The original boson-hopping term on a link
of the direct lattice is just a tight-binding hopping hamiltonian for
the ``line'' associated to the dual link crossing this direct link.  

The analogy to a tight-binding model thereby suggests a means of
removing the discrete constraint on $A_{a\alpha}$.  We can do this by
simply replacing the tight-binding model by a continuum one of a
particle in a periodic potential, strategically chosen so that the
lowest band is (arbitrarily) well-approximated by a tight-binding band,
well-separated from higher energy states.  This amounts to replacing
\begin{equation}
  \label{eq:band}
  -t \cos 2\pi E_{a\alpha} \rightarrow \frac{1}{2\kappa} E_{a\alpha}^2 - \tilde{t}
  \cos A_{a\alpha},
\end{equation}
and allowing $A_{a\alpha}$ to take arbitrary continuous real values.  The
tight-binding limit is recovered for $\tilde{t}\kappa \gg 1$.  One can then
adjust $\tilde{t},\kappa$ to achieve the desired tight-binding matrix
element $t$ between nearly-localized levels in neighboring wells of the
$\tilde{t}$ potential.  We do not do this explicitly here, since we will
not use it in the following.  It can, however, be achieved, by taking
$\tilde{t}, \kappa^{-1} \gg U$, to arbitary desired accuracy.
 
We are left with the hamiltonian
\begin{equation}
  \label{eq:Hdual3}
   \mathcal{H} = \sum_{a\alpha} \left[ \frac{1}{2\kappa} E_{a\alpha}^2 -
     \tilde{t} \cos A_{a\alpha}\right] + \tilde{U}
  \sum_i \left( (\epsilon_{\alpha\beta}\Delta_\alpha A_{\beta})_i-2\pi
    \overline{n} \right)^2,
\end{equation}
with $\tilde{U}=\frac{U}{(2\pi)^2}$.  This must be supplemented by the
commutation relations, Eq.~(\ref{eq:EAcom}), and the constraint,
Eq.~(\ref{eq:Gauss1}).  

Eq.~(\ref{eq:Hdual3}) is clearly a $U(1)$ gauge theory, but the presence
of ``vortex'' variables is not immediately apparent.  In fact, they are
implicit in the constraint, Eq.~(\ref{eq:Gauss1}), which can be
regarded as Gauss' law (in the dual
electromagnetic analogy).  Sites with non-zero $\Delta_\alpha
E_{a\alpha}$ thus correspond to dual charges -- physical vorticies.
Though no explicit vortex variables appear, they are un-necessary: the
locations of all (dual) charges can be determined from the electric
field lines.  

It is nevertheless useful (and conventional) to introduce redundant
vortex variables.  In particular, we introduce an auxiliary Hilbert
space of integer rotor states, eigenvectors of $N_a \in \mathcal{Z}$,
and conjugate variables $\theta_a$, with $[N_a,\theta_b]=i\delta_{ab}$
as usual.  These states are introduced only to redundantly label the
vortex positions.  So we require 
\begin{equation}
  \label{eq:Gauss2}
  \Delta_\alpha E_{a\alpha} = N_a.
\end{equation}
For this to be consistent with the hamiltonian action in the expanded
Hilbert space, we must ensure that the $-\tilde{t}\cos A_{a\mu}$ term,
which changes the electric divergence on neighboring sites, also
increment the new rotor variables.  This is accomplished by modifying
$A_{a\alpha} \rightarrow A_{a\alpha} - \Delta_\alpha\theta_a$.  Making
this shift, one sees that the hamiltonian in Eq.~(\ref{eq:Hdual3}) is
completely equivalent to the more conventional form
\begin{equation}
  \label{eq:Hdual4}
   \mathcal{H} = \sum_{a\mu} \left[ \frac{1}{2\kappa} E_{a\mu}^2 -
     \tilde{t} \cos (\Delta_\alpha\theta_a - A_{a\mu})\right] + \tilde{U}
  \sum_i \left( (\epsilon_{\mu\nu}\Delta_\mu A_{\nu})_i-2\pi \overline{n}\right)^2,
\end{equation}
combined with the constraint in Eq.~(\ref{eq:Gauss2}).

\subsection{Phases in the dual formulation, for integer $f$}
\label{sec:phas-dual-form}

For integer $\overline{n}$, we may remove the dual ``background flux''
$2\pi \overline{n}$ in the last term in Eq.~(\ref{eq:Hdual4}) by the
shift $A_{a2} \rightarrow A_{a2} + 2\pi f a_x$ (with $(a_x,a_y)$ the
coordinate of site $a$).  Here we are guaranteed $f=\overline{n}$ by
particle-hole symmetry of the on-site rotor model.  More generally,
working in the canonical ensemble with fixed filling $f$, this shift
by construction takes account exactly of the average dual flux in
$A_{a\alpha}$ -- the fluctuations around this shifted around are zero
by this choice.   One thus has just
a theory of lattice electromagnetism coupled to a ``charged scalar''
field $\psi_a = e^{-i\theta_a}$.  It may equally well be thought of as
a Ginzburg-Landau-like theory of a dual lattice ``superconductor''
with ``pair field'' $\psi_a$.

$ $From these analogies, one expects two phases.  In the gauge theory
language, there is a ``Coulomb phase'' or ``dielectric'', in
which the charged particle is gapped and can be integrated out.  A
simple limit of this phase is obtained by taking $\tilde{t}=0$.  In
this limit, $N_a$ is a constant of motion on each dual lattice site.
Because of the Gauss' law constraint, Eq.~(\ref{eq:Gauss2}), the
ground state is clearly obtained for $N_a=0$, i.e. no vortices
present.  Individual vortices can be introduced anywhere in the system
and in this limit have no dynamics, but due to the constraint, an
energy cost which is logarithmic in system size (due to the dual
electric field lines decaying as $1/r$ far from the vortex).  Clearly
this is the physical superfluid phase in the direct language.  As in
any superfluid, we expect a phason or Goldstone mode.  This
corresponds to the gapless linearly dispersing transverse photon of
the dual electromagnetism (there is only a single polarization in two
spatial dimensions).  Going back to the duality mapping, it is
straightforward to see that small $\tilde{t}$ corresponds to a large
``tight-binding bandwidth'' for $A_{a\alpha}$, i.e. a large
direct boson hopping amplitude, where we indeed expect a superfluid state.

The other phase is a ``Higgs phase'' or dual ``superconductor'', in
which the $\psi_a$ particle is condensed.  In this ``Higgs'' phase the
photon is gapped, and indeed there is an energy gap to all
excitations.  Moreover, there is no broken symmetry: the condensate
amplitude $\langle \psi_a \rangle$ is not gauge invariant and the
vortex condensation does not itself represent an order parameter for
any broken symmetry.  This then corresponds to the featureless Mott
insulating state in the direct picture.  

\subsection{Continuum field theory (for integer $f$)}
\label{sec:cont-field-theory}

For integer $f$, a rather na\"ive continuum limit is possible, due to
the ability to transform away the background ``flux'' in
Eq.~(\ref{eq:Hdual4}).  It is instructive, paralleling the logic used
above to derive the LGW theory coming from the Mott state, however, to
think more physically in terms of the vortex excitations.  We imagine
coming from the superfluid state, by increasing $\tilde{t}$ starting
from a small value. At first pass, we will neglect the fluctuations of
$A_{a\alpha},E_{a\alpha}$.  

For $\tilde{t}=0$, the ground state $|0\rangle$ clearly has $N_a=0$, and
corresponds to the vortex ``vacuum''.  The lowest excited single
vortex states correspond to $N_a=\pm 1$ on an (arbitrary) single site
$a$, which we denote 
\begin{eqnarray}
  \label{eq:vstates}
  |a+\rangle & = & \psi_a^\dagger |0\rangle = e^{i\theta_a}|0\rangle,
  \\
  |a-\rangle & = & \psi_a |0\rangle = e^{-i\theta_a}|0\rangle.
\end{eqnarray}
These states are elementary in that any vortex number configuration
can be built from them by superposition.  Each such state has a
logarithmically-divergent energy, since the Gauss' law constraint,
Eq.~(\ref{eq:Gauss2}), requires $E_{a\alpha} \sim 1/r$ far from the
vortex: this is just the usual logarithmic vortex energy coming from
the long-range superflow.  We will return to this point below.

First order degenerate perturbation theory in $\tilde{t}$ gives an
effective single-particle tight-binding model amongst the
$|a\pm\rangle$ states, and hence some plane-wave eigenstates with
energy dispersion.  One may now follow the same logic as in
Sec.~\ref{sec:integr-mott-superfl}, with the vortex/anti-vortex states
playing the analogous roles to the particle/hole excitations in the
Mott phase.  Time-reversal symmetry guarantees that these have the
same excitation gap.  Recognizing that these states carry the dual
gauge charge, and allowing for fluctuations in the
$A_{a\alpha},E_{a\alpha}$ fields, one thereby arrives (following the
same methodology as in Sec.~\ref{sec:integr-mott-superfl}, using the
coherent state path integral, and the standard path integral
Trotterization of electromagnetism) at the continuum action
\begin{equation}
  \label{eq:Sdualint}
  \mathcal{S}_{dual} = \int\!d\tau d^2x\, \big[
  \frac{1}{2e^2}(\epsilon_{\mu\nu\lambda} \partial_\nu A_\lambda)^2 +
  |(\partial_\mu 
    - i A_\mu)\varphi|^2 + \tilde{r} |\varphi|^2 + \tilde{u} |\varphi|^4+\cdots\big].
\end{equation}
Here for compactness, we have neglected to write unimportant
space-time anisotropies (constant scale factors) that appear between
space and time derivatives, and spatial (electric) and temporal
(magnetic) fluxes, writing simply three-dimensional coordinates
$(\tau,x,y)$ and indices $\mu,\nu,\lambda=0,1,2$.  The anisotropies do
not lead to any additional physics, and are expected to be irrelevant
or redundant at the QCP.  New coupling constants $\tilde{r},\tilde{u},
e^2$ have been defined in the continuum action.

Eq.~(\ref{eq:Sdualint}) is identical to the classical free energy of a
three-dimensional Ginzburg-Landau theory for a superconductor.  The
duality between this form and Eq.~(\ref{eq:Spsi}), the classical
three-dimensional XY free energy, was established in Ref.~\citen{dh}.  Because
the LGW form, Eq.~(\ref{eq:Spsi}), does not involve any gauge field,
it is of more practical (analytical and numerical) use in this
case.  Nevertheless, we emphasize that both the LGW and dual actions
are descriptions of the {\sl same} critical point.  Qualitatively the
same physics can be extracted from either form, though they will yield
different quantitative results from approximate treatments.  

It is instructive to review how the measures of charge localization
are encoded in Eq.~(\ref{eq:Sdualint}).  The boson creation operator,
$b_i^\dagger$, is a $2\pi$ {\sl flux insertion operator} in the dual
formulation, i.e. it creates a {\sl space-time monopole} with
$\partial_\mu B_\mu=2\pi \delta^{(3)}(x,\tau)$, where the dual
space-time magnetic flux $B_\mu = \epsilon_{\mu\nu\lambda}
\partial_\nu A_\lambda$ is physically the boson $3$-current.  By
looking at monopole correlators, therefore, one can discern the
presence or absence off-diagonal long-range order, or the presence or
absence of a gap.  The superfluid density $\rho_s$ and compressibility
$\kappa$ measure the response to external physical vector and scalar
potentials, respectively.  These physical potentials couple to the
boson $3$-current, and hence $\rho_s$ and $\kappa$ are obtained from
correlation functions of the $A_\mu$ gauge field.  The lesson to be
learned here is that {\sl all} Mott/Superfluid properties are encoded
in the properties of the gauge field $A_\mu$, not directly in the dual
``order parameter'' $\varphi$ (which anyway is not gauge-invariant --
See Sec.~\ref{sec:logar-vort-potent} for further related discussion).

A studious reader may wish to try the following illustrative exercise:
calculate the critical behavior of the superfluid density using the
dual form in the ``random phase approximation''.  This can be done by
introducing an external gauge field, $A_\mu^{\rm ext}$, which couples
to the physical electromagnetic current,
\begin{equation}
  \label{eq:extA}
  \mathcal{S} \rightarrow \mathcal{S} - i \int\! d\tau d^2x\, A_\mu^{\rm ext}
  \epsilon_{\mu\nu\lambda} \partial_\nu A_\lambda .
\end{equation}
Neglecting the $|\varphi|^4$ term, integrate out $\varphi$ at
quadratic level in $A_\mu$, then integrate out $A_\mu$ itself, to
obtain the coefficient of $(P_T A_\mu^{\rm ext})^2$ ($P_T$ is a
transverse projection operator), which is proportional to $\rho_s$.
One may compare the manner in which this vanishes with $\tilde{r}$
with the mean-field LGW prediction for how $\rho_s$ vanishes with $r$
in Eq.~(\ref{eq:Spsi}).

\subsubsection{Logarithmic vortex potential and bound states}
\label{sec:logar-vort-potent}

A careful reader may be troubled by the notion of using a vortex as an
elementary excitation, since, in the superfluid state, it has a
logarithmically divergent energy (in the system size).  A partial
answer to this question is that any neutral collection of equal
numbers of vortices and antivortices has finite energy.  However, it
is clear that a vortex and anti-vortex attract each other, and will
form a bound state, with infinite (logarithmically) binding energy.
On increasing $\tilde{t}$ within the superfluid state, one may suspect
that such a vortex/anti-vortex pair excitation will condense before
individual vortices do.  

In fact, more careful thought is needed.  In the ``relativistic''
theory of Eq.~(\ref{eq:Sdualint}) (or the lattice hamiltonian,
Eq.~(\ref{eq:Hdual4})) vortices and anti-vortices are not separately
conserved.  Indeed, the action of $\tilde{t}$ on the ground state
creates them in pairs on nearest-neighbor links (compare to the
$\lambda$ term in Eq.~(\ref{eq:Sph})).  Thus the number of such
``bound states'' is not a conserved quantity, and consequently there
is no sharp phase transition associated to their ``condensation'' --
any ``creation operator'' for these bound states has a non-vanishing
expectation value for all $\tilde{t}>0$, unless it is finely tuned.
Thinking spectrally, were such a single neutral bound state to approach zero
energy upon increasing $\tilde{t}$, it would be expected to have an
{\sl avoided crossing} with the ground state, because of the non-zero
matrix element between the na\"ive ground state and the state with one
excited bound pair.  

More generally, Eqs.~(\ref{eq:Sdualint},\ref{eq:Hdual4}) have {\sl no
  internal symmetries}, only space-time symmetries and ``gauge
invariance'' which is not a symmetry at all but embodies a dynamical
constraint.  Phase transitions in these models are then expected to be
either associated with the development of a Higgs mass for the gauge
field, or by breaking of spatial lattice symmetries (the latter not
being expected to occur for $f\in \mathcal{Z}$).

Nevertheless, it is a reasonable physical question to ask why it is
that a description in terms of elementary individual vortex
excitations is appropriate when these objects are always infinitely
strongly bound on the superfluid side of the transition.  The answer
is that, as the Superfluid-Mott QCP is approached, the interaction
between vortices is becoming progressively more and more ``screened''
by virtual fluctuations of vortex/anti-vortex pairs.  This can be
seen, for instance, from the fact that the superfluid density vanishes
at the QCP.  Alternatively, just from scaling, at the critical point,
the interaction between two static external dual ``test charges''
separated a distance $r$ behaves like $1/r$, not logarithmically.
A screened interaction prevails on scales smaller than the correlation
length, which diverges as the QCP is approached.  

\section{Non-integral Mott-Superfluid transitions: vortex theory}
\label{sec:non-integral-mott}

Having understood the direct and dual formulations of the integral
Superfluid-Mott transition, we turn to the non-integral case.  We will
concentrate on rational mean fillings, $f=p/q$, with $p,q$ relatively
prime, and $q>1$ but not too large.  We assume that sufficient
interactions are present to stabilize a ``crystalline'' Mott
insulating state at such a density.  We must then analyze
Eq.~(\ref{eq:Hdual4}), with $\overline{n} \approx p/q$
.\footnote{Strictly speaking, in the superfluid, the density is not
  exactly equal to $\overline{n}$, except when $\overline{n}$ is an
  integer or half integer in the microscopic rotor model.  What is
  important is not $\overline{n}$ but the density $f$ in the
  superfluid adjacent to the Mott state (within which $f$ is
  independent of $\overline{n}$ for a range of $\overline{n}$, it
  being incompressible).  The actual value of $\overline{n}$ should be adjusted
  to achieve $f=p/q$.  We tacitly assume this is done below.} \ 

If we attempt to approach the problem from an LGW perspective, there
is a fundamental difficulty: neither the Mott insulator or Superfluid
are ``disordered'' phases, the former breaking space-group symmetries
of the lattice, and the latter breaking the $U(1)$ boson conservation
symmetry.  Landau theory can only describe the transition from one of
these states to an even more symmetry-broken phase, which then seems
to require an intermediate phase between the two.

Taking a more physical point of view, we can search for pointlike
excitations in either phase that could provide the degrees of freedom
for a critical quantum field theory.  Unfortunately, on the Mott side,
the nature of the elementary excitations would seem to be very
specific to the particular Mott state under consideration.  All the
Mott phases presumably include extra/missing boson ``particles'', as
well as domain wall and other discrete topological excitations
particular to the precise type of boson density order in the ground
state.  The extra/missing boson ``particles'' could provide the
variables for some QCP, but one expects condensation of such particles
not to disrupt the density-wave order of the Mott state, leading to a
``supersolid'' rather than a true superfluid.  As such states appear
to be even more exotic than superfluids and Mott insulators, we will
not explore that possibility further.  A theory based on topological
excitations of particular Mott states is possible, but much more
limited in scope (see Ref.~\citen{psgbosons}, Sec.~?? for a
discussion).  

The most general approach seems to be from the superfluid.  This is
quite appealing insofar as, unlikely the panoply of possible Mott
states, the superfluid is unique, and has a symmetry which is
independent of the boson filling.  Thus the dual approach, based upon
vortex degrees of freedom, generalizes directly to arbitrary filling
factors.  Moreover, we will see that the variety of distinct Mott
states arises naturally from this description.

\subsection{Vortex projective symmetry group}
\label{sec:vort-proj-symm}

As for integral filling, we shift the dual vector potential to include
the average dual flux, $A_{a\alpha} \rightarrow
\overline{A}_{a\alpha}+A_{a\alpha}$, with 
\begin{equation}
  \label{eq:Abar}
  \overline{A}_{a2} = 2\pi f a_x, \qquad \overline{A}_{a1}=0,
\end{equation}
corresponding to Landau gauge.  Note that the specific gauge choice
(and indeed {\sl any} gauge choice) breaks na\"ive spatial lattice
symmetries.  Nevertheless, because
$\epsilon_{\alpha\beta}\Delta_\alpha\overline{A}_{a\beta}$ represents
a uniform gauge flux, this is an artifact of gauge fixing.  Clearly,
since $A_{a\alpha}$ is a dynamical variable, it is possible simply to
undo the effect of any space group operation on
$\overline{A}_{a\alpha}$ by an appropriate shift of $A_{a\alpha}$.
Because the gauge-invariant flux is not changed by such an operation,
this shift of $A_{a\alpha}$ must be pure gauge, i.e. a gradient
$\Delta_\alpha \chi_a$, for some scalar $\chi_a$.  As a consequence,
we can compose this shift with a pure gauge transformation -- of both
the vortex field $\theta_a$ (or equivalently $\psi_a$) and the gauge
field -- by the
phase factor $\chi_a$ in order to undo the shift of $A_{a\alpha}$.  By
this reasoning, it is clear that for every operation in the space-group,
there is a corresponding transformation consisting of a na\"ive
space-group transformation {\sl and} a gauge rotation of the vortex
fields {\sl without} any transformation of the gauge field.  This
transformation is almost unique, up to a global (i.e. $a$-independent)
$U(1)$ phase rotation of the vortex fields.  The global phase
arbitrariness is allowed because a uniform gauge rotation has zero
gradient and hence does not shift $A_{a\alpha}$.

In this way, by making some arbitrary phase choice for each group
element, one associates a transformation with each operation in the
space group.  These new transformations generate some new group, which
obeys the original group multiplication table {\sl up to phase
  factors}, which in general cannot be removed.  This is called a {\sl
  projective representation} of the space group, and in a slight abuse
of notation, we also call this new group a Projective Symmetry Group,
or PSG.  The proper subgroup (i.e. not including reflections) of the PSG
on the square lattice is generated by the three operations associated
with $x$ and $y$ translations, and a $\pi/2$ rotation (which we choose
about a dual lattice site), which can be chosen as:
\begin{eqnarray}
  \label{eq:psg1}
  T_y &:& \psi(a_x, a_y)  \rightarrow \psi (a_x, a_y - 1) \nonumber
  \\
  T_x &:& \psi(a_x, a_y)  \rightarrow \psi (a_x - 1, a_y) \omega^{
    a_y} \nonumber \\
  R_{\pi/2}^{\rm dual} &:& \psi(a_x, a_y)  \rightarrow \psi (a_y,
  -a_x ) \omega^{a_x a_y}, 
\end{eqnarray}
with $\omega \equiv e^{2\pi i f}$.  Specific algebraic relations follow from
these definitions, notably
\begin{equation}
  T_x T_y = \omega T_y T_x , \label{e7}
\end{equation}
which informs us that, unlike in the original space group, the
operations associated with translations in the PSG {\sl do not
  commute}.  The translational subgroup of the PSG is well-studied,
and Eq.~(\ref{e7}) is known as the magnetic translation algebra.  The
physical meaning of Eq.~(\ref{e7}) is, as explained in the
introduction, that the vortex acquires a dual Aharonov-Bohm phase of
$2\pi f$ upon encircling a site of the direct lattice, containing on
average $f$ bosons.  

The remaining algebraic relations between
generators are unchanged from the original space group:
\begin{eqnarray}
 T_x R_{\pi/2}^{\rm dual} &=& R^{\rm dual}_{\pi/2} T_y^{-1}  \nonumber \\
 T_y R_{\pi/2}^{\rm dual}
&=& R_{\pi/2}^{\rm dual} T_x \nonumber \\
\left( R_{\pi/2}^{\rm dual} \right)^4 &=& 1 \;. \label{rtr}
\end{eqnarray}

\subsection{Vortex multiplets} 
\label{sec:vort-quant-numb}

We again consider now the nature of the elementary vortex excitations,
starting from Eq.~(\ref{eq:Hdual4}).  As in
Sec.~\ref{sec:cont-field-theory}, at leading order in $\tilde{t}$, we
must solve the tight-binding model arising from degenerate
perturbation theory amongst the $|a\pm\rangle$ states, this time in
the presence of the mean gauge field, Eq.~(\ref{eq:Abar}).  This is
the Hofstadter problem, which is well-known generally to have a
extremely complex ``butterfly'' spectrum.\cite{hofstadter} We are,
fortunately, interested only in the lowest-energy states at the bottom
of the lowest Hofstadter band.  As for any particles, these must
appear in multiplets comprising an irreducible representation of the
symmetry group of the hamiltonian.  In this case, this group is the
PSG.

It is straightforward to see that, for a filling/dual flux with
denominator $q>1$, all representations of the PSG are {\sl at least}
$q$-dimensional.  This follows because one may choose to, say,
diagonalize $T_y$.  However, by Eq.~(\ref{e7}), acting with $T_x$ on a
state multiplies its eigenvalue of $T_y$ by $\omega$, hence this must
be a linearly independent state to the initial one.  This can be done
repeatedly $q-1$ times, until the $q^{th}$ time, one arrives back at a
state with the initial eigenvalue of $T_y$.  Thus the PSG connects
states of $q$ different values of the quasimomentum.  Since we used only
Eq.~(\ref{e7}), this conclusion is clearly true for the translational
subgroup of the PSG alone, so representations of the full PSG can only
be equal or larger.

It turns out, from explicit solution of the tight-binding model, that
a $q$-dimensional irreproducible representation (irrep) does exist for
the full PSG, and moreover, this is the lowest energy vortex multiplet
for the case of interest.  Details of this construction can be found
in Ref.~\citen{psgbosons}.  The result is that $q$ vortex fields,
$\varphi_\ell$, $\ell=0\ldots q-1$ may be defined, as relativistic
field operators (i.e. analogously to Eq.~(\ref{eq:lc}), as
superpositions of ``particle'' (vortex) creation and ``anti-particle''
(anti-vortex) annihilation operators) for each member of the
multiplet.  Under translations,
\begin{eqnarray}
T_x &:& \varphi_{\ell} \rightarrow \varphi_{\ell+1} \nonumber \\
T_y &:& \varphi_{\ell} \rightarrow \varphi_\ell \omega^{-\ell}.
\label{txty}
\end{eqnarray}
Here, and henceforth, the arithmetic of all indices of the
$\varphi_\ell$ fields is carried out modulo $q$, {\em e.g.\/}
$\varphi_q \equiv \varphi_0$.  Under the rotation,
\begin{equation}
 R_{\pi/2}^{\rm dual} : \varphi_{\ell} \rightarrow \frac{1}{\sqrt{q}}
\sum_{m=0}^{q-1} \varphi_m \omega^{-m \ell}, \label{rpi}
\end{equation}
which is a Fourier transform in the space of the $q$ fields.  The full
(improper) PSG is generated by including also reflections (defined
here about $x$ and $y$ axes of the dual lattice),
\begin{eqnarray}
I_x^{\rm dual} &:& \varphi_\ell \rightarrow \varphi_{\ell}^{\ast}
\nonumber
\\
I_y^{\rm dual} &:& \varphi_\ell \rightarrow \varphi_{-\ell}^{\ast} \;.
\label{ixiy}
\end{eqnarray}

A continuum field theory can be now constructed from the
$q$-dimensional vortex multiplet.  The effective action should be
invariant under all the physical symmetries, with vortex fields
transforming under the PSG and the dual $U(1)$ gauge symmetry.
Restoring fluctuations of the gauge field, the critical action
constructed in this manner, generalizing Eq.~(\ref{eq:Sdualint}) to
non-integer filling, has the form given in Eq.~(\ref{eq:Sq}) in the
introduction. 

\subsection{Order parameters}
\label{sec:order-parameters}

It is important to consider the observables in the problem.  For
integer filling, we discussed that $\varphi$ is not itself a true
``order parameter'' because it is not gauge invariant.  Physical
quantities (e.g. superfluid density, compressibility, off-diagonal
long-range-order) in that case are related just to properties of the
dual gauge field.  This is because local, gauge-invariant combinations
of the single vortex field such as $|\varphi|^2$ etc. (related to the
vortex/anti-vortex bound states discussed in
Sec.~\ref{sec:logar-vort-potent}) are scalars under all symmetries of
the hamiltonian.  

The situation is dramatically different for $q>1$.  While the above
physical quantities are still related to properties of the dual gauge
field, in general, gauge-invariant bilinears of the form
$\varphi_\ell^* \varphi_{\ell'}^{\vphantom{*}}$ are not scalars under
the spatial symmetries.  They can thus serve as order parameters for
various types of symmetry breaking.  Group theoretically, the direct
product of an irrep of the PSG and its conjugate can be decomposed
into a sum of true irreps {\sl of the ordinary space group} (not
projective representations).  Happily, the basic components of these
irreps can be constructed in generality.\cite{psgbosons} \ With the
definition of Eq.~(\ref{e11}) of the introduction, straightforward
manipulations show that $\rho({\bf Q})$ transforms exactly as expected
for a Fourier component of a scalar ``density'' with wavevector ${\bf
  Q}$.  Specifically, it is easy to verify that $\rho_{mn}^{\ast} =
\rho_{-m,-n}$, and from Eqs.~(\ref{txty}) and (\ref{rpi}) that the
space group operations act on $\rho_{mn}$ just as expected for a
density wave order parameter
\begin{eqnarray}
T_x &:& \rho_{mn} \rightarrow \omega^{-m} \rho_{mn} \nonumber
 \\
T_y &:& \rho_{mn} \rightarrow \omega^{-n} \rho_{mn} \nonumber
 \\
R_{\pi/2}^{\rm dual} &:& \rho_{mn} \rightarrow \rho_{-n,m}.
\label{e12}
\end{eqnarray}

\section{Examples}
\label{sec:examples}

While the set of different $\rho_{mn}$ can describe a variety of
different density wave orders in the Mott state, the number of such
ordering patterns is limited, so the vortex theory actually constrains
the nature of Mott insulating density ordering occuring in the
neighboring of a continuous transition to a superfluid.  These orders
can be determined from a mean-field analysis of the effective action
$\mathcal{S}$, Eq.~(\ref{eq:Sdualint}).  We give examples for
$q=2,3$.  Further examples can be found in Ref.~\citen{psgbosons}.

\subsection{Half-filling and deconfined criticality}

The case of $q=2$, corresponding to bosons at half-filling, is
particularly interesting.  One has ${\cal L}_{\rm int} = {\cal L}_4 +
O(\varphi^6)$, with
\begin{eqnarray}
  \mathcal{L}_4 = \frac{\gamma_{00}}{4} \left( |\varphi_0|^2 +
    |\varphi_1|^2 \right)^2 + \frac{\gamma_{01}}{4} \left( \varphi_0
    \varphi_1^\ast - \varphi_0^\ast \varphi_1 \right)^2. \label{m2}
\end{eqnarray}
It is convenient to make the change of variables
\begin{eqnarray}
  \varphi_0 &=& \frac{\zeta_0 + \zeta_1}{\sqrt{2}} \nonumber \\
  \varphi_1 &=& -i \frac{\zeta_0 - \zeta_1}{\sqrt{2}}. \label{m3}
\end{eqnarray}
The action in Eq.~(\ref{m2}) reduces to
\begin{eqnarray}
\mathcal{L}_4 = \frac{\gamma_{00}}{4} \left( |\zeta_0|^2 +
|\zeta_1|^2 \right)^2 - \frac{\gamma_{01}}{4} \left( |\zeta_0|^2 -
| \zeta_1|^2 \right)^2. \label{m4}
\end{eqnarray}
The result in Eq.~(\ref{m4}) is identical to that found in earlier
studies \cite{lfs,sp} of the $q=2$ case.

Minimizing the action implied by Eq.~(\ref{m4}), it is evident
that for $\gamma_{01} < 0$ there is a one parameter family of
gauge-invariant solutions in which the relative phase of $\zeta_0$
and $\zeta_1$ remains undetermined. As shown in earlier
work \cite{lfs}, this phase is pinned  at specific values only by an
$8^{th}$ order term proportional to $\sim (\zeta_0 \zeta_1^\ast )^4 +
\mbox{c.c}$.

The mean-field analysis finds three phases:
First, phase (A): an ordinary charge density wave (CDW) at
wavevector ($\pi,\pi$). The other two states are VBS states in
which all the sites of the direct lattice remain equivalent, and
the VBS order appears in the (B) columnar dimer or (C) plaquette
pattern. The phases (A), (B), and (C) appear in the upper-left,
upper-right, and lower-right corner of the inset in Fig.~\ref{fig:2}.
Note that the ``site-centered stripe'' phase in the lower-left of the
inset, though it certainly can occur in lattice boson models, does not
occur according to the vortex theory in the vicinity of the transition
to the superfluid.  The saddle point values of the fields associated with
these states are:
\begin{eqnarray}
(A)~&:&~ \zeta_0 \neq 0~,~\zeta_1 = 0~\mbox{or}~\zeta_0
= 0~,~\zeta_1 \neq 0. \nonumber \\
(B)~&:&~ \zeta_0 = e^{i n \pi/2} \zeta_1 \neq 0. \nonumber \\
(C)~&:&~ \zeta_0 = e^{i (n+1/2) \pi/2} \zeta_1 \neq 0, \label{m5}
\end{eqnarray}
where $n$ is any integer. 

\subsubsection{Deconfined criticality}
\label{sec:deconf-crit}

This particular example, for $\gamma_{01}<0$, has recently been
understood in much more detail.  It turns out that this case,
describing the transition from a superfluid to a columnar or plaquette
VBS state, can be understood from a complementary point of view as a
theory of {\sl fractional} ``half''-boson excitations.  It is too
involved to fully explore this in detail in this paper, but we will
at least uncover these fractional excitations in the dual theory.

For $\gamma_{01}<0$, the mean-field solutions in the Mott state have
both $\zeta_\ell=|\zeta|e^{i\vartheta_\ell}$, with $|\zeta|$ constant.
Supposing slowly
varying $\vartheta_\ell$, the effective phase-only action is
\begin{equation}
  \label{eq:rhos}
  S_{eff} = \int\! d^2r d\tau\, \sum_\ell \frac{\rho'_s}{2}
  |\partial_\mu \vartheta_\ell - A_\mu|^2 + \frac{1}{2e^2}
  (\epsilon_{\mu\nu\lambda}\partial_\nu A_\lambda)^2,
\end{equation}
with $\rho'_s = 2 |\zeta|^2$.  Consider a fixed `vortex'
(independent of $\tau$) in one -- say $\vartheta_0$ -- of the $2$
phase fields, centered at the origin $r=0$.  One has the spatial
gradient $\vec\nabla\vartheta_0 = \hat\phi/r$, while
$\vec\vartheta_1=0$ (here $\hat\phi =
(-y,x)/r$ is the tangential unit vector) .  Clearly, the action is
minimized for tangential $\vec{A} = A \hat\phi$.  Far from the
`vortex' core, the Maxwell term
$(\epsilon_{\mu\nu\lambda}\partial_\nu A_\lambda)^2$ is
negligible, so one need minimize only the first term in
Eq.~\ref{eq:rhos}.  The corresponding Lagrange density at a
distance $r$ from the origin is thus
\begin{equation}
  \label{eq:vortexlag}
  {\cal L}'_v = \frac{\rho'_s}{2} \left[(1/r-A)^2 + A^2\right].
\end{equation}
Minimizing this over $A$, one finds $A=1/(2r)$.  Integrating this
to find the flux gives
\begin{equation}
  \label{eq:fluxprime}
  \oint \vec{A}\cdot d\vec{r} = \pi
\end{equation}
Since the physical charge is just this dual flux divided by
$2\pi$, the `vortex' in $\zeta_0$ indeed carries fractional
boson charge $1/2$.  

Note that, because of the small $8^{th}$ order term locking the phases
$\vartheta_0$ and $\vartheta_1$ together, such a 'vortex' in just one
of these fields costs a divergent energy (actually linear in system
size).  This indicates these fractional particles are ``confined'' in
the Mott state, like quarks in quantum chromodynamics.  This
confinement, however, becomes weaker and weaker as the SF-Mott QCP is
approached, due to the smallness of the $8^{th}$ order term in this
limit.  In this sense -- and in others discussed in Refs.~\cite{dqcp}
-- this is a ``deconfined QCP''.

A na\"ive extension of this argument would suggest the presence of
charge $1/q$ for general $q$.  It turns out that this generalization
is not so simple, and while it can be made in some cases, there are
considerable restrictions involved -- see Ref.~\citen{psgbosons}.

\subsection{$q=3$}
\label{sec:q3}

\begin{wrapfigure}[3]{r}{6.6cm}
\centerline{\includegraphics[width=2.7in]{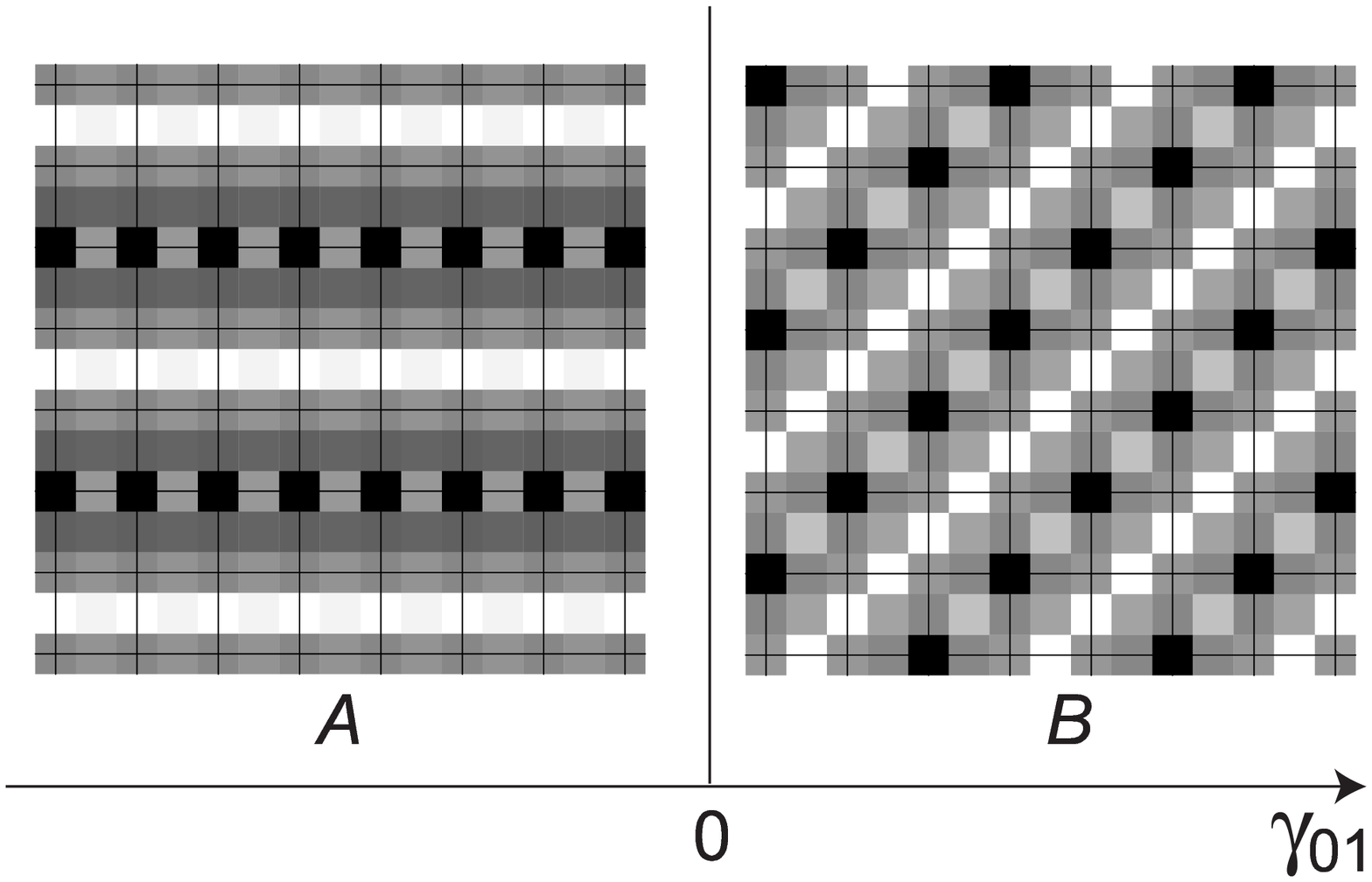}}
\caption{Charge-ordering patterns for $q=3$}
\label{figq3}
\end{wrapfigure}

Now there are 3 $\varphi_\ell$ fields, and the quartic potential
in Eq.~(\ref{m2}) is replaced by
\begin{eqnarray}
\mathcal{L}_4 &=& \frac{\gamma_{00}}{4} \left( |\varphi_0|^2 +
|\varphi_1|^2  + |\varphi_2|^2 \right)^2 \nonumber \\ &+&
\frac{\gamma_{01}}{2} \left( \varphi_0^\ast \varphi_1^\ast
\varphi_2^2 + \varphi_1^\ast \varphi_2^\ast \varphi_0^2 +
\varphi_2^\ast \varphi_0^\ast \varphi_ - \varphi_0^\ast
\varphi_1^2 + \mbox{c.c.} \right. \nonumber \\
&-& \left. 2 |\varphi_0|^2 |\varphi_1|^2 - 2 |\varphi_1|^2
|\varphi_2|^2- 2 |\varphi_2|^2 |\varphi_0|^2\right). \label{m6}
\end{eqnarray}
The results of a mean-field analysis for this potential are shown in
Fig~\ref{figq3}.

The states have stripe order, one along the diagonals, and the
other along the principle axes of the square lattice. Both states
are 6-fold degenerate, and the characteristic saddle point values
of the fields are
\begin{eqnarray}
(A)~&:&~ \varphi_0 \neq 0~,~\varphi_1 = \varphi_2 =
0~\mbox{or}\nonumber \\
&~&~~~e^{i 4n \pi/3}\varphi_2 = e^{i 2 n \pi/3} \varphi_1
=
\varphi_0 \neq 0 \nonumber \\
(B)~&:&~ \varphi_0=\varphi_1 = e^{\pm 2 i \pi/3}
\varphi_2~\nonumber \\
&~&~~~\mbox{and permutations}. \label{m7}
\end{eqnarray}

\section*{Acknowledgements}
We thank M.~P.~A.~Fisher and T.~Senthil for valuable discussions. This
research was supported by the National Science Foundation under grants
DMR-9985255 (L.  Balents), DMR-0098226 (S.S.), and DMR-0210790,
PHY-9907949 at the Kavli Institute for Theoretical Physics (S.S.), the
Packard Foundation (L. Balents), the Deutsche Forschungsgemeinschaft
under grant BA 2263/1-1 (L. Bartosch), and the John Simon Guggenheim
Memorial Foundation (S.S.). S.S. thanks the Aspen Center of Physics
for hospitality. K.S. thanks S.~M.~Girvin for support through ARO
grant 1015164.2.J00113.627012.


%

\end{document}